# NMR based Pharmaco-metabolomics: An efficient and agile tool for therapeutic evaluation of Traditional Herbal Medicines

Dinesh Kumar,[1,*] Atul Rawat,[1,2] Durgesh Dubey,[1,2] Umesh Kumar,[1] Amit K Keshari,[3] Sudipta Saha[3] and Anupam Guleria[1,*]

[1]Centre of Biomedical Research, SGPGIMS Campus, Raibareli Road, Lucknow-226014
Department of [2]Biotechnology and [3]Pharmaceutical Sciences, Babasaheb Bhimrao Ambedkar University, Vidya Vihar, Rai Bareli Road, Lucknow 226025

*Authors for Correspondence:
Dr. Dinesh Kumar: dineshcbmr@gmail.com
Dr. Anupam Guleria: anupam@cbmr.res.in

**Abstract:**

Traditional Indian (Ayurvedic) and Chinese herbal medicines have been used in the treatment of a variety of diseases for thousands of years because of their natural origin and lesser side effects. However, the safety and efficacy data (including dose and quality parameters) on most of these traditional medicines are far from sufficient to meet the criteria needed to support their world-wide therapeutic use. Also, the mechanistic understanding of most of these herbal medicines is still lacking due to their complex components which further limits their wider application and acceptance. Metabolomics -a novel approach to reveal altered metabolism (biochemical effects) produced in response to a disease or its therapeutic intervention- has huge potential to assess the pharmacology and toxicology of traditional herbal medicines (THMs). Therefore, it is gradually becoming a mutually complementary technique to genomics, transcriptomics and proteomics for therapeutic evaluation of pharmaceutical products (including THMs); the approach is so called pharmaco-metabolomics. The whole paradigm is based on its ability to provide metabolic signatures to confirm the diseased condition and then to use the concentration profiles of these biomarkers to assess the therapeutic response. Nuclear magnetic resonance (NMR) spectroscopy coupled with multivariate data analysis is currently the method of choice for pharmaco-metabolomics studies owing to its unbiased, non-destructive nature and minimal sample preparation requirement. In recent past, dozens of NMR based pharmaco-metabolomic studies have been devoted to prove the therapeutic efficacy/safety and to explore the underlying mechanisms of THMs, with promising results. The current perspective article summarizes various such studies in addition to describing the technical and conceptual aspects involved in NMR based pharmaco-metabolomics.

## Introduction:

Metabolomics -a newborn cousin to genomics and proteomics- is an analytical approach to metabolism that involves quantitative and comparative analysis of concentration profiles of low molecular weight metabolites and their intermediates in affected biological systems (typically urine, blood-plasma/serum, cell lysates, or tissue extracts). Genomics, transcriptomics and proteomics analysis complemented with metabolomics information, offers the potential to understand the whole biological system including health or disease processes operating in that system –so called systems biology approach **(Fig. 1).** With its ability to discover disease related biomarkers and underlying biochemical processes, today, metabolomics is used virtually in all aspects of biomedical research aiming to improve the understanding of the health and disease processes. The whole paradigm is based on the fact that a pathophysiological condition or therapeutic intervention results in a specific and characteristic change in the biochemical composition profiles of biofluids and metabolomics aims to identify these changes. The biochemical changes -that correlate to a disease (or disease type/grade) and/or treatment response- then allow the clinical researchers to improve diagnosis and treatment of disease including, early disease detection, monitoring response to treatment and patient stratification for treatment. The molecular biomarkers validated on wide-range of human populations form the basis for new clinical diagnostic assays.

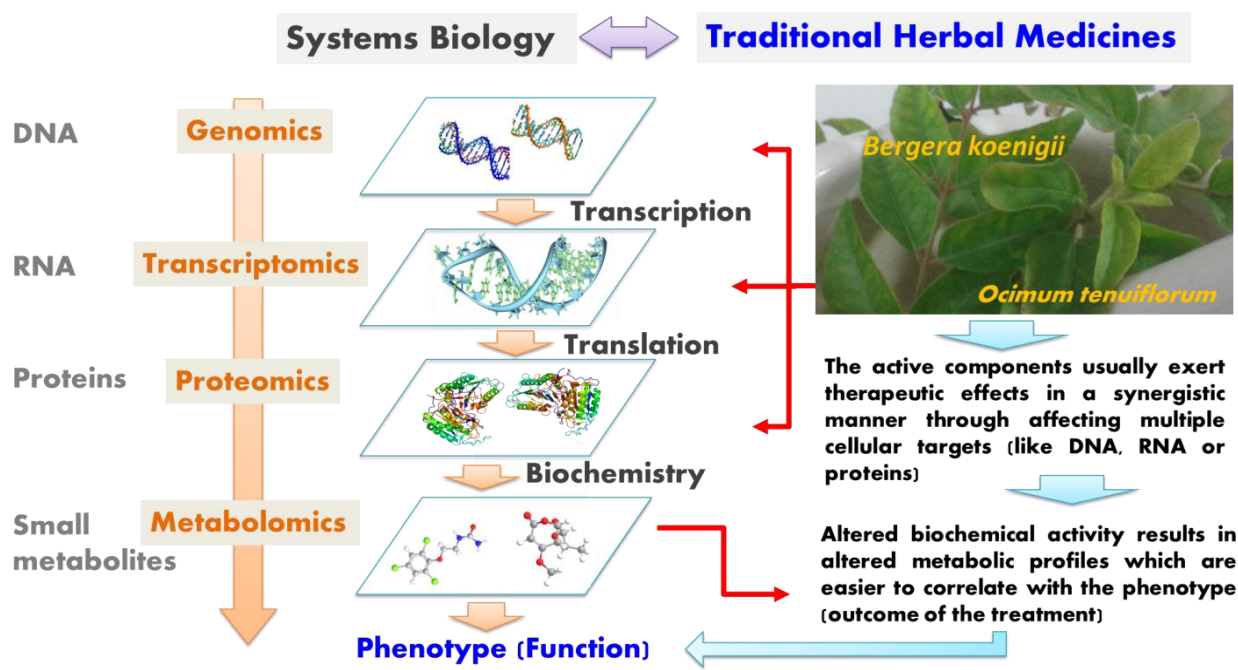

**Figure 1:** Schematic showing (on left) technological platforms of systems biology approach: genomics, transcriptomics, proteomics, and metabolomics and (on right) generalized mode of action of traditional herbal medicines (THMs). Overall, it highlights the fact that the active components of THMs can interact with various cellular components (i.e. DNA, RNA, and proteins) to exert combined synergistic therapeutic effect.

Metabolomics is also gaining tremendous popularity in pharmaceutical research owing to close technical and conceptual overlap between the two[1]. In particular, because of its ability to reveal altered metabolism in response to a therapeutic intervention, it is now routinely used in pharmaceutical research[2] particularly meant for (a) understanding the underlying mechanisms of drug action, (b) studying the biochemical effects produced in response to a drug treatment and (c) assuring the quality, safety and efficacy of pharmaceutical products. Such an explosion of metabolomics applications in various important areas of pharmaceutical research has led to origin of new discipline called **Pharmaco-metabolomics**[3,4] which particularly aims to improve the efficacy and reduce the side effects associated with a treatment. Pharmaco-metabolomics is also widely used as complementary tool in understanding biological changes after a drug intervention or a gene knockout[5]. Recently, it is gaining tremendous popularity in ethano-pharmacological studies aiming to evaluate the efficacy and safety of traditional herbal medicines (THMs) [1,3-6].

**Table 1:** Summarized $^1$H NMR based pharmacometabolomics studies carried out to investigate the pharmacological outcome of traditional herbal medicines:

| # | Traditional Herbal Medicine | Evaluated drug effect on rat/mouse models | Ref |
|---|---|---|---|
| 1 | Curcumin[1] | Evaluated anti-Hyperlipidemic properties | Li Z Y et. al.[7] |
| 2 | Sini decoction[2] | Evaluated therapeutic efficay against myocardial infarction | Tan G et. al. [8] |
| 3 | *Coptidis Rhizome*[2] | Evaluated drug-induced gastrointestinal reaction | Zhou Y et. al.[9] |
| 4 | Niuhuang Jiedu Tablet (NJT)[2] | Evaluated its $As_2S_2$ toxicity alleviation effect | Xu W et. al.[10] |
| 5 | Herba Rhodiolae[2] | Evaluated its Anti-hypoxia and anti-anxiety effects | Liu X et. al.[11] |
| 6 | Astragali Radix (Huangqi)[2] | Evaluated its anti-fatigue effect | Li Z Y et. al.[12] |
| 7 | Danggui (DG)[2] | Evaluated and compared the drug efficacy of Chinese and European variants of Danggui | Zhang Z Z et.al.l[13] |
| 8 | Xiaoyaosan[2] | Evaluated its therapeutic response in depressed patients | Liu C C et. al. [14] |
| 9 | Astragali Radix[2] | Compared the clinical efficacy and safety of two different forms of Astragali Radix | Li A P et. al. [15] |
| 10 | Qin-Re-Jie-Du (QRJD) and Liang-Xue-Huo-Xue (LXHX)[2] | Evaluated differential efficacy of these traditional formulations in septic rats. | Li Y et. al.[16] |
| 11 | Curcuma aromatica oil[1] | Evaluated its intervening effects on renal interstitial fibrosis rats | Zhao L et. al.[17,18] |
| 12 | Centella asiatica[1] | Evaluated its *in vivo* antidiabetic property on obese diabetes rats | Abas F et. al.[19] |
| 13 | Total Alkaloid of Corydalis Rhizoma[2] | Evaluated its therapeutic efficacy against depression on chronic unpredictable mild stress rat model | Hongwei W et. al.[20] |

[1]**Traditional Herbal Medicine;** [2]**Chinese Herbal Medicine;**

The most information-rich techniques currently employed in metabolomics studies are mass spectrometry (MS) and nuclear magnetic resonance (NMR) spectroscopy. Among these two techniques, MS is highly sensitive, with detection limits in the picogram range, however, for MS-based metabolomics studies, it requires a separation step (liquid chromatography/gas chromatography) before the MS detection. This requirement dictates that each sample will requires hours per MS analysis. The overall throughput is further hampered by many unsolved problems such as (a) non-uniform detection caused by variable ionization efficiency, (b) lack of well-established and standardized methods or procedures (as it requires optimization of separation conditions each time), and (c) the difficulties still met in the identification of novel/unknown metabolites produced in response to a given therapeutic intervention. Although less sensitive (with detection limits in the low micromolar to sub-micromolar range) than MS, NMR is generally appreciated for metabolic analysis because of its versatility for analyzing metabolites in the liquid state (serum, urine, and so on), in intact tissues (for example, tumors) or *in vivo* (e.g. brain, liver, kidney, and so on). The powerful strengths of NMR spectroscopy are reproducibility, the ability to quantitate compounds in unmodified complex biological mixtures and the ability to identify unknown metabolites[21-23]. In recent past, the dozens of NMR based pharmaco-metabolomic studies have been carried out to prove the therapeutic efficacy/safety and to explore the underlying mechanisms of THMs, with promising results. **Table 1** summarizes some of these reports where the $^1$H NMR-based metabonomics approach has been used to investigate the pharmacological outcomes with traditional herbal medicines. The metabolomics approach has also been employed for studying the adverse toxic effects on experimental models (in terms of perturbed biochemical composition profiles) in response to traditional herbal medicines)[24,25].

## Technological Advantages of NMR:

Compared to other analytical and biochemical analysis methods used for metabolomics studies, NMR offers several clear advantages[26]: (a) first, it is applicable for a variety of biological and clinical samples, tissue extracts and even cell lysates, (b) second, it is rapid, quantitative, and offers the potential for high-throughput (i.e. analysis of >100 samples/day is attainable), (c) third, it is least-destructive (i.e. the prepared sample can be used in multiple consecutive NMR experiments, or after the NMR experiments are completed, the sample can be analyzed by other analytical techniques), (d) fourth, it is un-biased (i.e. all protonated metabolites present in a biological mixture are detectable irrespective of their physical properties) and (e) last but not the least, it requires virtually no sample preparation and provides highly reproducible results. These are the reasons that NMR has become the method of choice for studying metabolic alterations associated with distinct human pathologies and also gaining tremendous popularity in pharmaco-metabolomics studies as well[27]. Today, the sensitivity of NMR is also not a major issue; even low nanogram detection limits are possible with appropriate instrumentation and novel pulse methodology[28]. Further, the recent technological advances in cryogenically cooled probe technology, the miniaturizing of sample probe head (i.e. with microvolume probe it is now possible to analyze samples in 10 µl solution), higher field-strength superconducting magnets and high-performance radiofrequency coils have increased the sensitivity of NMR spectrometers by a factor of ~ 3–4[29,30].

## General Guidelines for NMR based Pharmaco-metabolomics:

A typical NMR based pharmaco-metabolomics study employed to evaluate the therapeutic efficacy and safety of THMs is performed in two steps: (a) first step involves the identification of the disease specific biomarkers –referred here as **disease metabotyping (Fig. 2)** and (b) second step involves the use of these biomarkers to assess the therapeutic efficacy. The various technical and conceptual aspects related to both these steps are described further.

## (A) Disease metabotyping:

Disease metabotyping involves identification of disease specific metabolic profiles referred here as metabolic signatures or biomarkers. The identified metabolic signatures then allow the drug researchers and drug regulators to address many of the riskier or expensive issues associated with the discovery, development and monitoring of novel therapeutic candidates through providing early preclinical indications of efficacy and toxicity of a potential drug candidate[26]. **Figure 2** schematically illustrates the generalized protocol employed in NMR based disease metabotyping. Typically, it involves the following steps:

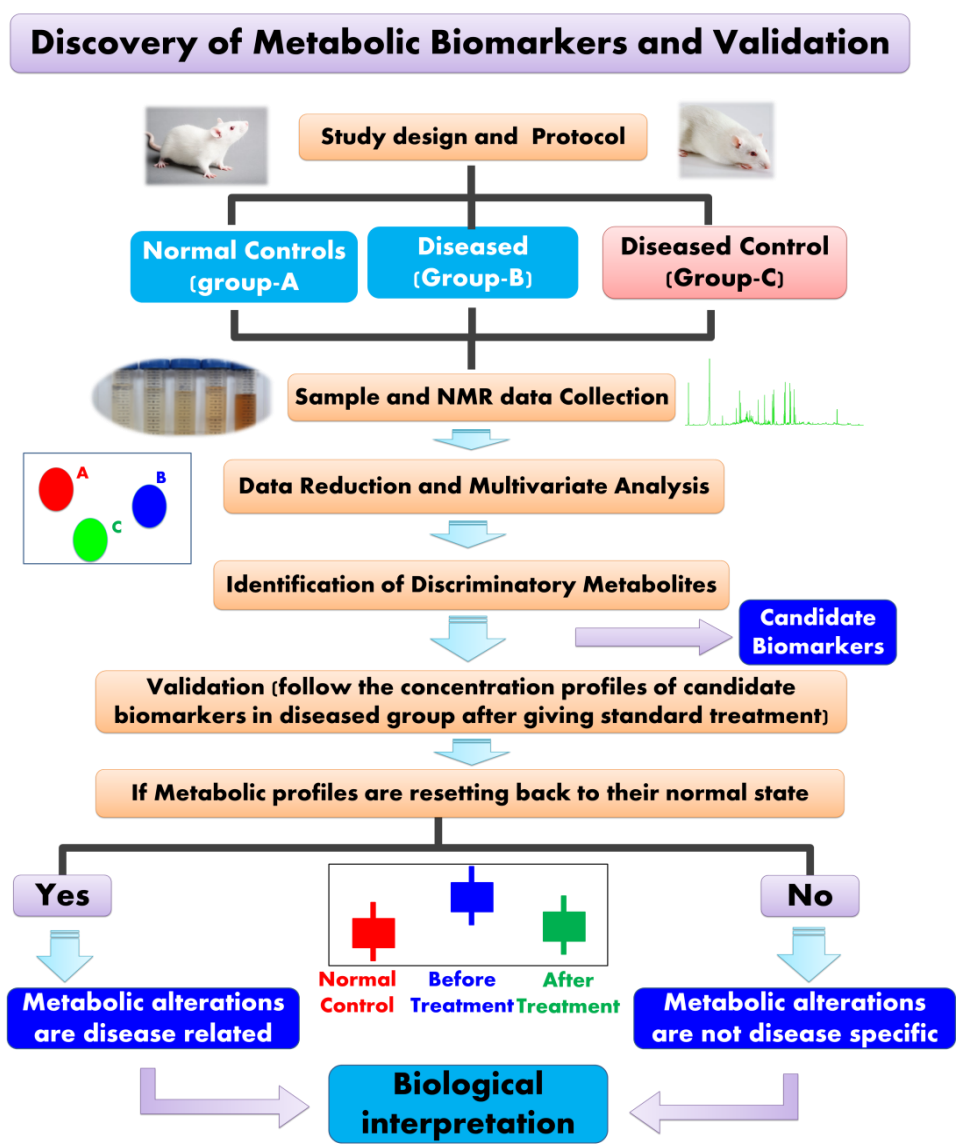

**Figure 2:** Flowchart depicting generalized protocol for identifying disease-specific metabolic signatures using NMR based metabolomics approach.

### (a) Animals experiments

Typically, the NMR-based pharmaco-metabonomics studies focus on the metabolic profiles of biofluids (specifically urine, serum and tissue extracts) obtained from normal and treated experimental animal models. As such, these studies involve male rats/mice which are first subjected to acclimatization (at least for seven days). Throughout acclimatization and intervention period, the rats/mice are provided with uniform diet and tap water ad libitum and are generally kept under ambient temperature of 25 ± 2 °C with 12 hours day and night cycle and 40% - 60% relative humidity. After acclimatization, the rats/mice are divided randomly into a control group and a few (generally 3-4) intervention groups. For toxicological studies, the normal rats are given different doses of the treatment. Whereas for pharmacological studies, all the rats/mice in each group are first induced with diseased condition or then given the therapeutic treatment as per the study protocol. At the end of treatment period, rats/mice from each group are subjected to ethological evaluation following the protocols described previously[31,32]. After ethological evaluation, all experimental rats/mice are sacrificed by decapitation and required blood and tissue samples are quickly extracted from each rat/mouse, snap-frozen in liquid nitrogen and subsequently stored at -80 °C before NMR analysis. For urinary metabolic profiling, urine samples are collected at regular time intervals before and during the intervention period. Generally midstream urine is collected (minimum 0.5 ml) and transferred to appropriately labeled tubes, immediately cooled or flash-freezed, and stored at -80 °C. For extracting serum samples, the whole blood is collected in a plain tube and incubated at room temperature (generally 25 °C) for 30 minutes. As a general rule, 1.0 ml of whole blood yields ~0.5 ml of serum which is well sufficient for NMR data collection in conventional 5 mm tubes. The resulted supernatant is then transferred to a 1.5 ml microcentrifuge tube (MCT) and centrifuged at 2,500 rpm for 10 minutes (at room temperature to precipitate the sediments). The resulted supernatant – referred as serum- is then transferred to appropriately labelled sterile tubes, immediately cooled or flash freezed, and stored at -80 °C. For metabolomics studies, the stored samples are thawed 1-2 hours before starting the NMR experiments and prepared following the previously reported NMR sample preparation protocols[33,34]. Depending upon the situation in hand or availability of the biofluid, the users may change the protocol. Such an optimal protocol for urine and serum sample preparation for NMR based metabolomics studies is depicted here schematically in **Figure 3**. Regarding NMR of tissue samples, it has other specific requirements depending upon whether the metabolic profiles of tissue samples are studied (a) directly i.e. using high-resolution magic angle spinning (HR-MAS) NMR as described previously[35] or (b) indirectly i.e. tissue extracts are prepared [34,36,37] and characterized using standard solution state NMR. It is important to mention here is that the NMR chemical shifts are highly sensitive to sample conditions (like sample temperature, pH, ionic strength, etc.), therefore, while collecting urine/blood, it is important to treat all of the samples the same way to avoid inter-spectral variation of chemical shifts. Though, urine samples contain minimal fat and protein content and therefore owing to its non-viscous nature, the NMR spectra of urine **(Figure 4A)** show the highest sensitivity and resolution. However, inter-spectral variations are common particularly in case of urine samples. This is because, the biochemical composition profiles of urine are influenced by a range of different factors e.g. diet, time of food intake, amount of water consumption, seasonal variations, gender of the subject (male/female) or physical activity prior to sample collection[38]. Further, collecting urine samples from animal models is an arduous job and requires utmost precaution to avoid contamination. On the other hand, the biochemical composition profiles in sera extracted from blood collected after overnight fasting are less sensitive for such variations rendering serum samples quite apposite for routine metabolomics studies.

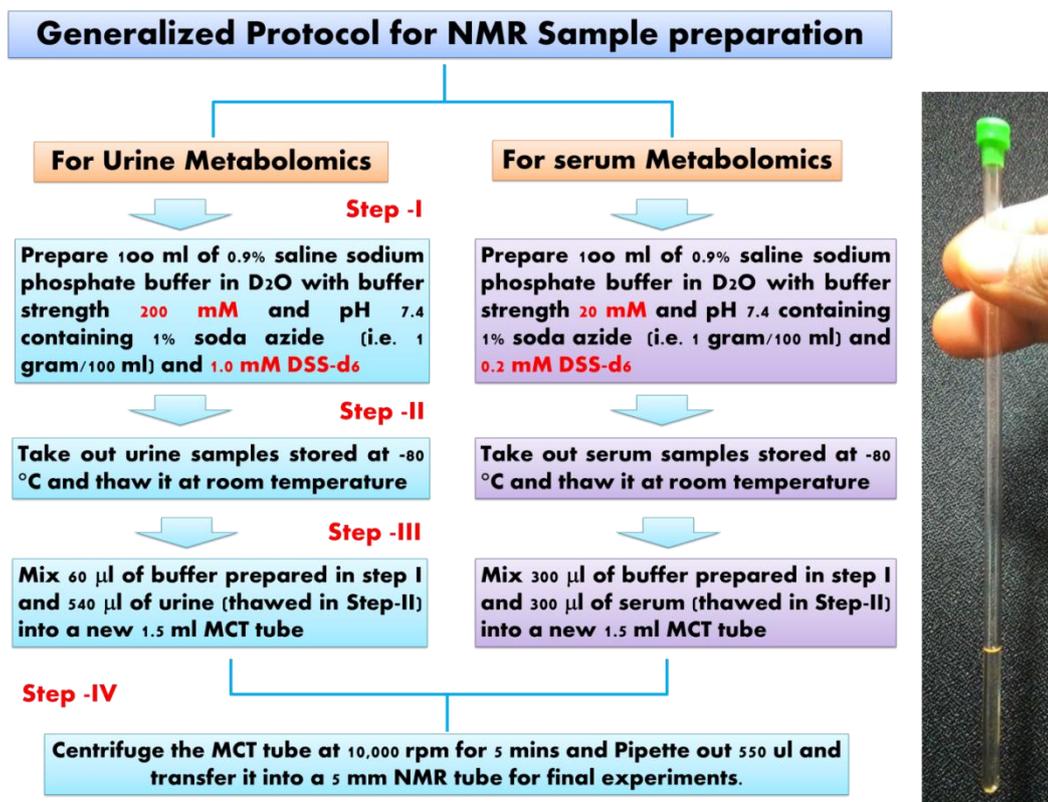

**Figure 3:** Flowchart showing generalized protocol for preparing the urine and serum samples for NMR based metabolomics studies. The protocol assumes that in each case, more than 300 μl of serum and more than 540 μl of urine is obtained.

**(b) NMR experiments:**

One-dimensional (1D) $^1$H NMR is widely used technique for quantitative proilfing of complex metabolite mixtures of bio-fluids (like urine and serum) and average time spent for sample preparation and NMR data collection is generally 10-15 minutes. Routinely, after data collection, the metabolite signals observed in the $^1$H NMR spectra of serum/urine are assigned (as depicted in **Figure 4B**) by comparison with spectra of standard compounds (www.bml-nmr.org) or performing metabolite spiking with standrad authentic compounds[39]. Additional 2-dimensional NMR experiments (including J-resolved, $^1$H-$^1$H TOCSY and heteronuclear $^1$H-$^{13}$C HSQC correlation experiments) are performed for the purpose of confirming chemical shift assignments using software programs like Metabominer[40,41]. In addition, the metabolite peaks can be assigned making use of chemical shift databases (www.hmdb.ca), previosuly reported chemical shifts in the literature, and software tools like CHENOMX (Edmonton, Canada))[39,41].

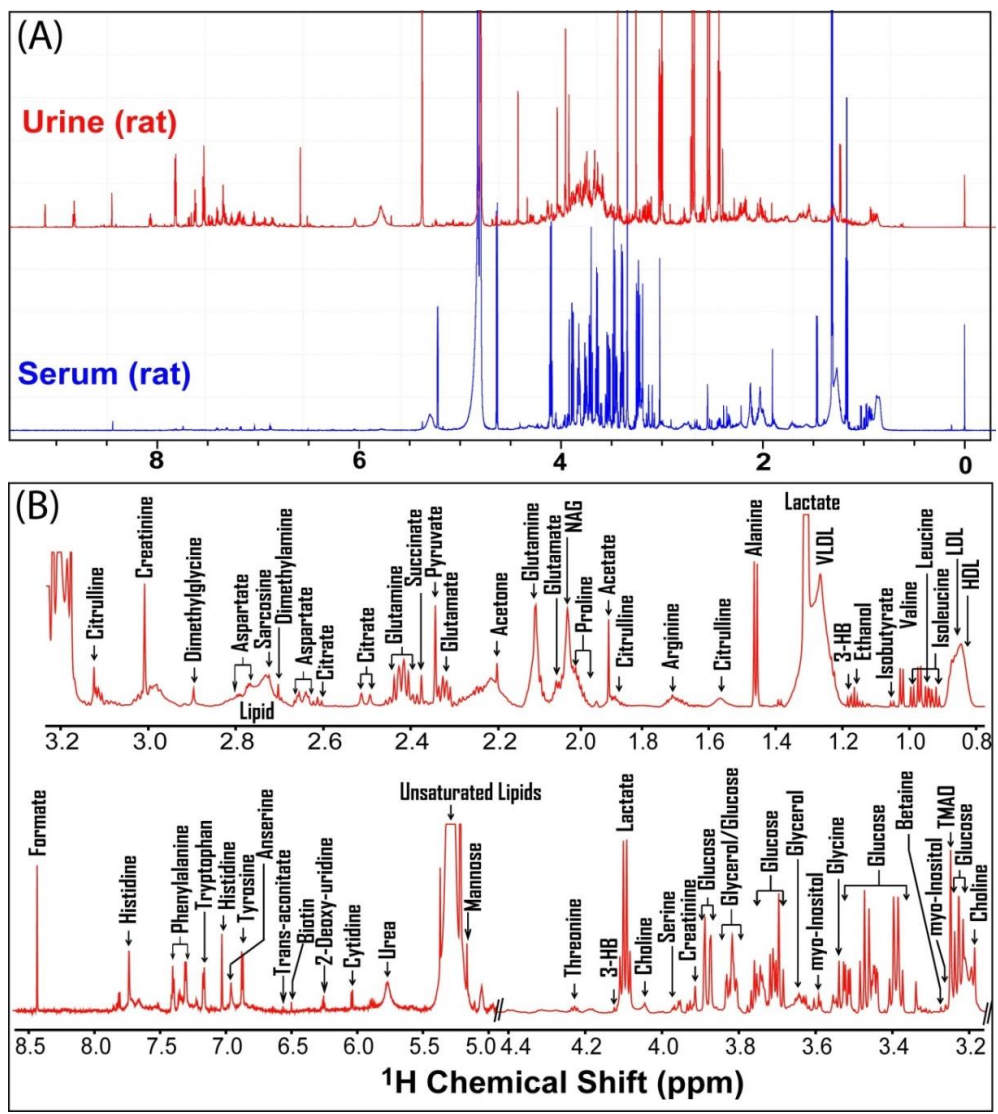

**Figure 4: (A)** Stack plot of representative 800 MHz one-dimensional $^1$H CPMG (Carr-Purcell-Meiboom-Gill) NMR spectra of rat urine **(in red)** and rat serum **(in blue)** recorded using cpmgpr1d pulse sequence from Bruker library with water presaturation during recycle delay of 5 sec and T2 filter time of 60 ms (for suppressing the singals of higher MW proteins and fats). Unlike urine, the 1D $^1$H NMR spectrum of serum is relatively simple and less crowded as evident in **(B)**. The lables in **(B)** represent the metabolite specific assignment of peaks using in CHENOMX profiler.

### (c) Statistical Analysis:

Following NMR data collection and spectral assignment, the next step is to identify the distinct metabolic patterns or disturbances associated with the diseased state. Owing to spectral complexity, various data refinement and multivariate statistical analysis tools are used to extract the relevant information[42]. First, the acquired series of NMR spectra are processed (including phase and baseline correction) and calibrated generally with respect to an internal reference (typically lactate in case of serum and creatinine in case of urine). Next, the spectral data is transformed into a data matrix following various data pre-processing and refinement steps: (a) binning i.e. each spectrum is divided into equally sized (typically 0.01-0.04 ppm) bins **(Fig. 5a)**, so that integral (area) of each bin represents a new point in the binned spectrum, (b) normalization where each spectrum is normalized such that its integral is 1 (this is so called internal scaling and is used to account for variable dilution factors of biological samples), and (c) centering and scaling (a table column operation; is used to reduce the influence of intense peaks while emphasizing weaker peaks)[43-45]. The software tools available to perform various such operations include AMIX (Bruker Biospin), CHENOMX NMR suite (www.chenomx.com) and Mnova NMR (http://mestrelab.com).

After binning, normalization and scaling, the resulted data matrix is subjected to multivariate statistical analysis simulatenously to reduce the dimensionality of the data and recognizing the trends/patterns[46-52]. Typically, it includes two steps: unsupervised and/or supervised analysis. Unsupervised analysis aims at finding intrinsic variation within a group (i.e. grouping trend) and patterns in the data using methods such as hierarchical clustering analysis (HCA, often used to assess, how similar or different the diseased sample groups are compared to normal control samples)[50,52] and principal component analysis (PCA, a data reduction technique used to reduce the dimensionality of a multi-dimensional dataset while retaining the characteristics of the dataset that contribute most to its variance)[45,46,53].

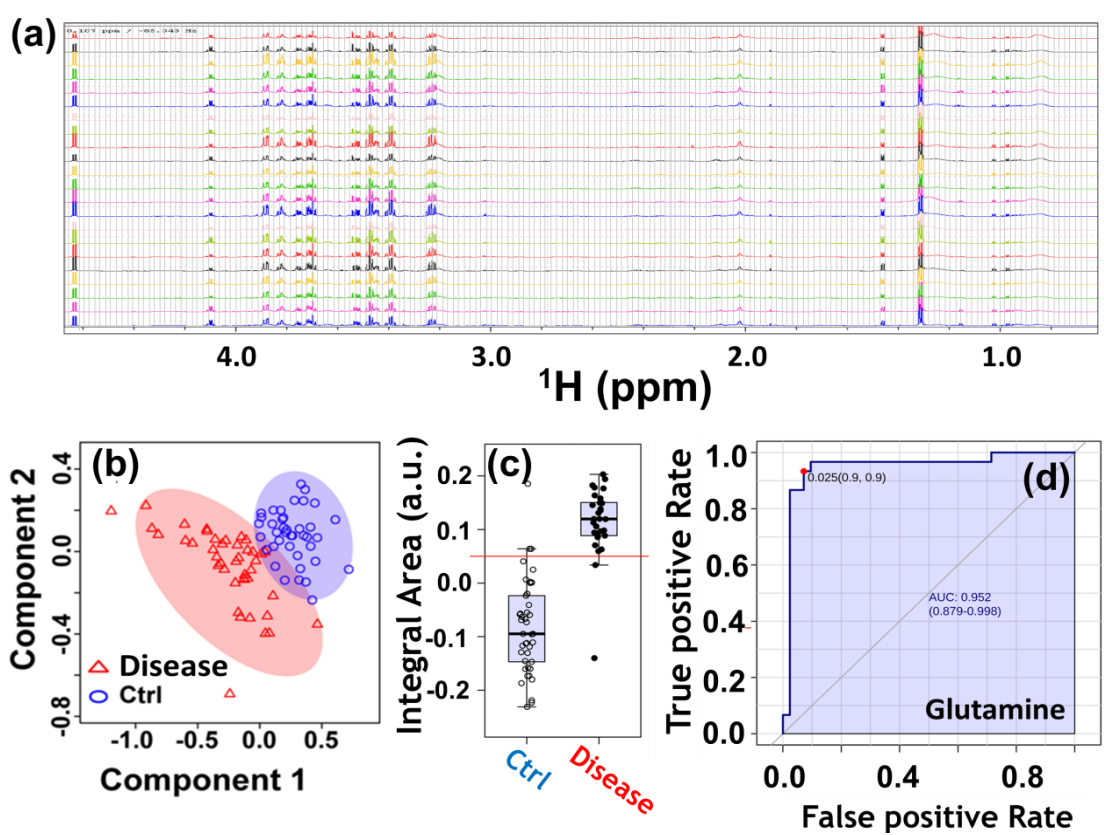

**Figure 5:** (a) Representative arrayed NMR dataset recroded on rat serum samples (related to an ongoing pharmaco-metabolomics study in the lab) with vertical grey lines highlighting the data binning (or so called bucketing). (b) An illustrattive PLS-DA score plot (with semi-transparent confidence intervals) derived from NMR based serum metabolic profiles of control and diseased groups. (c) A representative box-cum-whisker plot of a example discriminatory metabolite (i.e. putative biomarker) elevated in diseased state. Horizontal line inside the box is the median and bottom and top boundaries of boxes are 25th and 75th percentiles, respectively. Lower and upper whiskers are 5th and 95th percentiles, respectively, and (d) corresponding ROC (Receiver operating characteristic) curve plot based on the serum levels observed in control and diseased state highlighting the diagnostic potential of the biomarker.

Supervised analysis, on the other hand, allows for statistical inference regarding features and groups to assess the clustering behavior between groups. The most common statistical model applied for supervised analysis is Partial Least Squares for Discriminant Analysis (PLS-DA) which is a regression extension of PCA and takes advantage of class information to attempt to maximize the separation between groups of observations. The PLS-DA analysis also helps to identify the marker metabolites that can differentiate the two groups. These metabolites are further verfied using univariate analysis including student t-test, ROC (Receiver operating characteristic) curve analysis **(Fig. 5b-d)**. However, supervised analysis based on PLS-DA model often tends to over fit the data and therefore the model needs to be rigorously validated to see whether the separation is statistically significant or is due to random noise[54]. The commonly used software tools for multivariate data analysis include MUMA-R_Package[55], MVAPACK[56], Unscrambler (CAMO USA, Norway: www.camo.com), SIMCA (Soft Independent modeling by Class Analogy: umetrics.com/products/simca), Metaboanalyst (http://www.metaboanalyst.ca)[50,52], etc. Important to mention here is that the Metaboanalyst is a free web-based server which allows comprehensive metabolomic data analysis, visualization and validation [52]. The example PLS-DA score plot shown in **Figure 5b** has been generated using Metaboanalyst server. In addition, metaboanalyst also perfoms model validation (R-squared analysis), helps to evaluate model quality (Q-squared analysis) and does allow the univariate analysis (including box plot data visualization and ROC curve analysis). The box plot (showing difference in the concentration profiles of a discriminatory metabolite identified based on PLS-DA model) and the corresponding ROC-curve shown in **Figure 5d** and **5e** have also been generated using Metaboanalyst server. On top of all this, the other strength of metaboanalyst is that it further allows integrative pathway analysis of genes and metabolites for biologically relevant interpretion of observed biochemical changes (as depicted in **Figure 6**); thus helps to validate and generate disease hypothesis for scientific acceptance[52]. Overall, the NMR spectroscopy coupled with multivariate statistical analysis allows the identification of metabolic disturbances and/or specific metabolic pathways associated with the disease, but it also allows the identification of metabolic signatures which have their potential diagnostic and prognostic implications for clinical management of the disease.

**(B) Assessing Therapeutic efficacy and safety of traditional Herbal Medicines (THMs) using metabolomics analysis**

Therapeutic evaluation of herbal drugs is one of the essential steps employed for demonstrating their utility in clinical practice. However unlike a conventional medicine, a herbal formula based on a THM is composed of complicated chemical components which work as a holistic system for curing the disease. Biologically, a THM produces the therapeutic effect synergistically affecting multiple cellular targets. Therefore, for in-depth understanding of underlying efficacies and mechanisms of action for such holistic herbal recipes, the systems biology has been recognized as a novel approach[57-59]. Its technological platforms, such as genomics, transcriptomics, proteomics and metabolomics, together provide in-depth understanding of whole range of biological and biochemical effects (including molecular interactions, molecular pathways and regulatory networks) **(Figure 1)**. The systems biology approach is also gaining tremendous popularity for assessing the pharmacological and toxicological effects of herbal recipes.

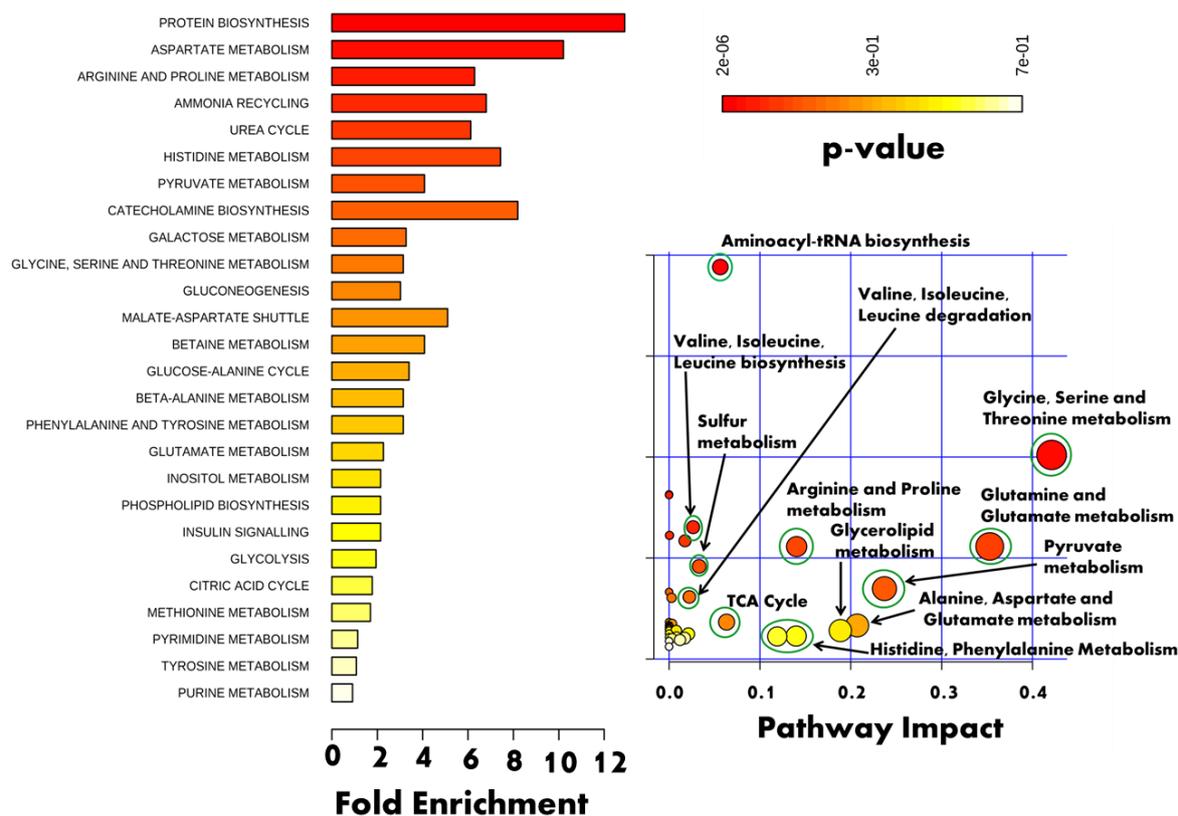

**Figure 6:** Summary of metabolic pathways enrichment analysis performed in MetaboAnalyst (Version 3.0, URL: http://www.metaboanalyst.ca) using a set of 21 metabolites found to be significantly altered in AMI patients compared to normal controls (in a preliminary study from our lab). The used metabolite entities were: Isoleucine, Leucine, Valine, Alanine, Acetate, Glutamate, Glutamine, Citrate, TMAO, Glucose, Glycine, Glycerol, Creatine, Creatinine, Serine, Histidine, Phenylalanine, Lactate, Proline, Threonine, and Choline. The most significant p-values are in red while the least significant are in yellow and white. Pathway analysis showing altered metabolic pathways.

Metabonomics is a top-down systems biology approach **(Figure 1)** and plays a key role in the convergence of systems biology studies. Today, it is also extensively used in basic and applied research for assessing the pharmaceutical efficacy and toxicity. The $^1$H NMR spectroscopy coupled with multivariate data analysis is currently the technique of choice for metabolomics studies meant for studying the biochemical composition profiles and metabolic pathway perturbations from urine and blood samples. **Figure 7** schemically illustartes the generalized protocol employed in NMR based pharamcometabolomics studies and **Figure 8** schemically shows the metabolic reprogramming after therapeutic intervention (in dose dependent manner). Additionally, the data collected for evaluating the therapeutic efficacy of the drug can be used to widen the understanding of metabolic pathways implicated in response to drug treatment and therefore can help unravel the underlying mechanism of drug action **(Fig. 5)**. Additional health related issues associated with THMs include the quality control of the production and optimum dosage. Metabolomics, however, can also be applied for the assessment of quality control of herbal medicines and finding optimum non-toxic dose of such THMs as elegantly described previously[58-62].

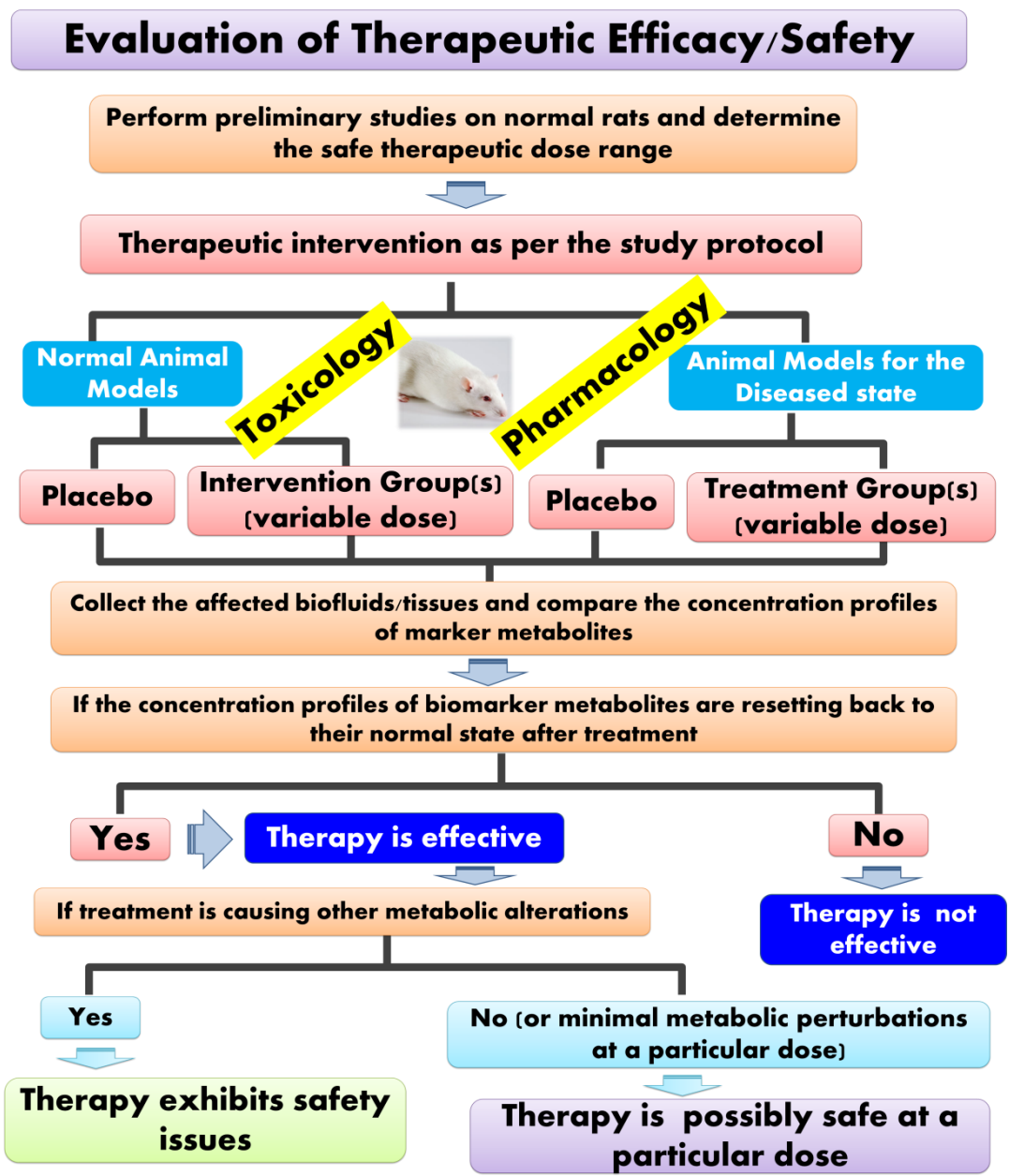

**Figure 7:** Flowchart depicting generalized protocol for evaluating the *in vivo* efficacy and safety of new therapeutic formulation using metabolomics approach.

**Concluding Remarks:**

Pharmaco-metabolomics is a rapidly emerging discipline in biomedical research pertaining to human-healthcare and providing a new perspective for studying the outcome of a treatment including herbal medicines. The central to pharmaco-metabolomics is to characterize the metabolic-perturbations/metabolic-pathways implicated in response to a therapeutic intervention. It further offers a useful tool to identify disease biomarkers and therefore provides a novel methodological cue for systematically dissecting the underlying mechanisms of drug action and biochemical effects produced in response to drug treatment. In this report, we have highlighted the potential of NMR for screening pharmacological outcomes. The particular advantage of the approach is its rapidity and reproduciblity with little or no added technical resources to existing *in vivo* studies. New technological and methodological advances in combination with rapid generation of metabolomic databases and automated algorithms for data processing and analysis has significantly improved the qualitative and quantitative power of this approach rendering it a useful tool in pharmaceutical research and development including mechanistic exploration. A future challenge, however, is to validate the pharamco-metabolomics findings in large and prospective, well-controlled clinical studies involing demographically diverse human patient populations. Nevertheless, the approach holds a great promise to widen the understanding of underlying efficacies and mechanisms of action.

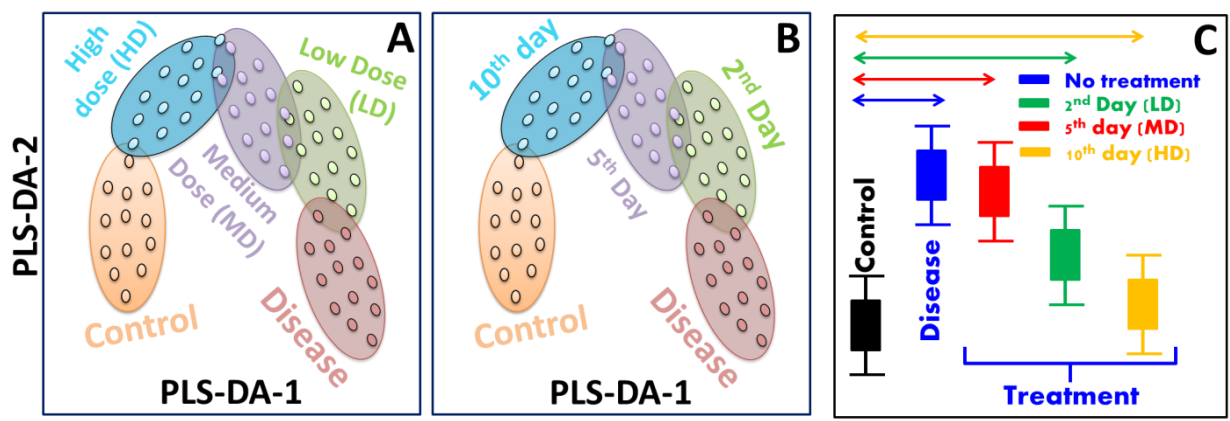

**Figure 8:** Schematic PLS-DA score plots depicting metabolic reprogramming duing treatment **(A)** or after treatment in a dose dependent manner **(B)**. **(C)** box plot of a metabolite elevated in diseased state and resetting back to its normal level after treatment.